\documentstyle[pra,epsf,aps]{revtex}

\def\cm#1{}

 \def\Kappa{K}

\begin{document}
\title{{
Variational Resummation for $ \epsilon $-Expansions
of Critical Exponents of
Nonlinear  O($n$)-Symmetric $ \sigma $-Model
in $2+ \epsilon $ Dimensions}}
\author{Hagen Kleinert%
 \thanks{Email: kleinert@physik.fu-berlin.de \hfil \newline URL:
http://www.physik.fu-berlin.de/\~{}kleinert \hfil
}}
\address{Institut f\"ur Theoretische Physik,\\
Freie Universit\"at Berlin, Arnimallee 14,
14195 Berlin, Germany}
\maketitle
\begin{abstract}
We develop a method for extracting
accurate critical exponents from perturbation
expansions of the O($n$)-symmetric nonlinear $ \sigma $-model in $D=2+ \epsilon $ dimensions.
This is possible by considering the $ \epsilon$-expansions
in this model as strong-coupling expansions of functions of
the variable $\tilde  \varepsilon \equiv  2(4-D)/(D-2)$, whose first five
weak-coupling expansion coefficients of powers of $\tilde  \varepsilon $
are known from $  \varepsilon $-expansions of critical exponents in
O($n$)-symmetric $\phi^4$-theory in $D=4- \varepsilon$ dimensions.
\end{abstract}

%
~\\
\noindent
{\bf 1.}
Critical exponents
of the O($n$)-universality class
can be calculated
with high accuracy from standard resummation
procedures of renormalization group functions of
$\phi^4$-field theory
\cite{gz,MN}.
For the classical
Heisenberg model, where $n=3$,
the critical exponent $ \nu $ governing the divergence
of the coherence length as $\xi\propto |T-T_c|^{-\nu}$
has been calculated
from seven-loop perturbation expansions
in three dimensions \cite{MN}
as $ \nu =0.7073\pm 0.0030$ \cite{gz},
whereas  five-loop expansions in $D=4- \varepsilon $ dimensions
\cite{ksf} extrapolated to $ \varepsilon =1$
give
 $ \nu =0.7050\pm 0.0055$ \cite{gz}.
Apart from the initial expansion coefficients, the resummation procedures
incorporate
information on the large-order growth of the coefficients
obtained from semiclassical considerations \cite{Z,KS}.
Results very close to the above numbers were recently obtained from a novel
strong-coupling
$\phi^4$-theory \cite{SC,seven}
in $D=3$ \cite{seven} as well as
$4- \varepsilon$ dimensions \cite{k263,ksf2}.

It is generally accepted that,
as a consequence of the {\em universality hypothesis}
of critical phenomena of all systems with equal Goldstone bosons,
the same critical exponents should be obtainable
from renormalization group studies
of O($n$)-symmetric nonlinear $ \sigma $-models
in $D=2+ \epsilon $ dimensions at $ \epsilon =1$,
if the second-order character of the transition
is not destroyed by fluctuations.
These conditions restrict the comparison to $n>2$.
For $n=1$ (Ising case), there are no Goldstone bosons,
and for $n=2$ (XY-model), the transition is of infinite order,
for which the divergence of the correlation length with
temperature cannot be parametrized
like $\xi\propto |T-T_c|^{-\nu}$,
as shown by Kosterlitz and Thouless.

Unfortunately, the $ \epsilon $-expansions
of the nonlinear $ \sigma $-models
have, up to now, remained rather useless
for any practical calculation,
due to their non-Borel character \cite{HB}.
This has led some authors
to doubt the use of such expansions around the lower critical dimension
altogether \cite{cc}.
Basis of these doubts is the
increasing relevance of ignored
higher powers of the derivative term in the calculations \cite{weg}.
Such a situation would be quite unfortunate, since
it would jeopardize other interesting
theories which depend on similar relationships, such as
Anderson's theory of localization \cite{And}.

Fortunately,
the counter-arguments are not completely convincing
since they involve
an interchange of limits
in the analytic continuation in $ \epsilon $ and
the increase of the number of derivatives
\cite{BH}, so that hope remains.
The purpose of this note is to confirm this hope and
to lend further
support to the intimate relationship
of $ \epsilon $- and $ \varepsilon $-expansions.
This is done by
determining
from a combination of the two expansions
an accurate
critical exponent $ \nu $ for the classical Heisenberg model
for all dimensions $2\le D\le4$.

\begin{figure}[tbhp]
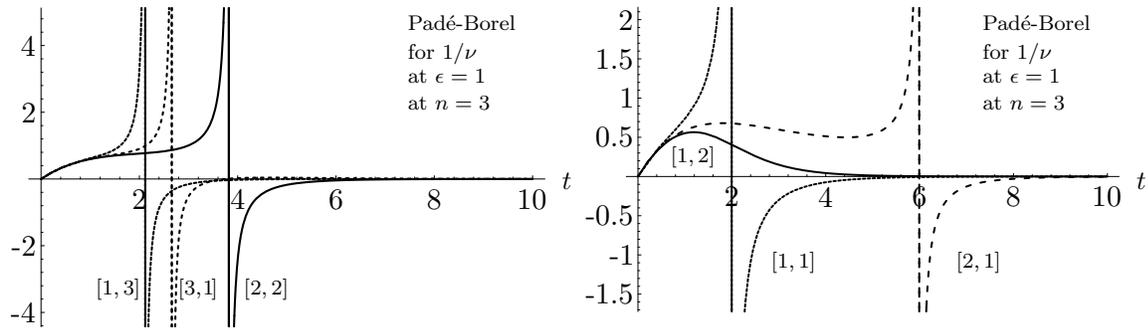

\hspace{-3.5cm}\input pade4.tps
\hspace{3.5cm}\input pade.tps   ~\\
\caption[]{
Integrands of the Pad\'e-Borel
transform (\ref{@pb})
for the Pad\'e approximants
[1,3], [3,1], [2,2] and for
[1,1], [2,1], [1,2]
at $ \epsilon =1$.
Only the last is integrable, yielding $ \nu ^{-1}\approx 1.25183\approx1/.79883$.
}
\label{pade}\end{figure}

~\\
\noindent
{\bf 2.}
So far, the
 $ \epsilon $-expansions of $ \nu ^{-1}$ and the anomalous dimension $ \eta $
have been calculated up to the powers $ \epsilon ^4$ \cite{HB,new}:
\begin{eqnarray}
 \nu ^{-1}( \epsilon )&=& \epsilon+ \frac{\epsilon ^2}
	   {n-2} + \frac{\epsilon ^3}{2(n-2)}
-\left[30-14+n^2 +(54-18n)\zeta(3)\right] \frac{\epsilon ^4}{4(n-2)^3}
+ \dots~,\label{@ps}               \\
 \eta( \epsilon ) &=&
 { \epsilon \over {n-2}  }  -{{\left(  n-1 \right) \epsilon ^2 }
\over
    {{{\left( n-2 \right) }^2}}}  +
  {{}}{n\left( n -1\right)\epsilon ^3
\over {{{2\left( n-2 \right) }^3}}}
-
  {{\left( n -1\right) \,\left[ -6 + 2\,n + {n^2} +(- 12
+
        n+ {n^2})\,{\rm \zeta}(3) \right] }}{ \epsilon ^4
\over
    {4\,{{\left( n-2 \right) }^4}}}+\dots~ .
\label{@ps2}\end{eqnarray}
The singularity at $n=2$ reflects
the above-discussed restriction of the upcoming considerations to $n>2$.
When evaluated at $ \epsilon =1$,
the first series yields for the three-dimensional
O(3)-model the
diverging successive values $ \nu ^{-1}=(1,\,2,\,2.5\,,3.25)$.
The often-employed Pad\'e approximations do not help,
with the best of them, the [1,2]-approximation, giving
the too large value $ \nu =2$.
So far, the only
result which is not too far from the
true value
has been obtained via
the
Pad\'e-Borel transform  \cite{HB}
\begin{equation}
P^{[1,2]}( \epsilon ,t)=\frac{ \epsilon t}{1- \epsilon t/2+ \epsilon ^2t^2/6},
\label{@}\end{equation}
from which one obtains
the $ \epsilon $-dependent inverse critical exponent
\begin{equation}
 \nu ^{-1}( \epsilon )=
\int _0^\infty dt\,e^{-t}P^{[1,2]}( \epsilon ,t).
\label{@pb}\end{equation}
Its value  at $ \epsilon =1$ is $\nu ^{-1}\approx1.252$, corresponding to
$ \nu \approx0.799$, which is still considerably
larger
than the accurate value $0.705$.
The other Pad\'e-Borel approximants are singular
and thus of no use. See Fig.~\ref{pade} for plots of the integrands.


A direct evaluation of the series for the other critical exponent, the
anomalous dimension
$ \eta $,  yields
the successive  values $(2,\,-2,\,4,\,-5)$, which are completely useless.
Here the nonsingular Borel-Pad\'e approximations
$[2,1],\, [1,2]$, and $[1,1]$
yield $0.147$, $0.150$, and $0.139$,
rather than the correct value $0.032$.

~\\
\noindent
{\bf 3.}
The remedy for these problems
comes from a combination of the
theory developed in Refs.~\cite{SC,seven} with a procedure
developed in Ref.~\cite{int}.
The theory allows us
to extract the
strong-coupling  properties of a $\phi^4$-theory
from perturbation expansions.
In particular, it renders the
power behavior of
the renormalization constants
for large bare couplings $g_0$, and from this all critical exponents
 of
the system.
By using the known expansion coefficients of the
renormalization constants
in three dimensions up to six loops,
we were able to derive extremely accurate
values for the critical exponents.
The method is a systematic
extension
 to arbitrary orders \cite{systematic}
of the Feynman-Kleinert variational approximation
to path
 integrals \cite{tk}.
For
an anharmonic oscillator,
this so-called
{\em variational perturbation theory\/} \cite{PI}
yields expansions
 which converge  uniformly
in the coupling strength
and exponentially fast, like $e^{-{\rm const}\times N^{1/3}}$
in
 the order
$N$
of
   the approximation, as was observed in \cite{JK1,JK2,PI}
and
proved in \cite{JK3,Guida}.
The extension to field theory was achieved in Ref.~\cite{SC},
and showed
to same type of convergence,
but with
the fractional power $1/3$
replaced by the irrational power $1- \omega$, where  $\omega$
is
the critical exponent governing the approach to
scaling.

This theory is combined with
the procedure of Ref.~\cite{int}
which allows us to
interpolate
variationally functions for which we
know strong- and weak-coupling
expansions.
The resummation to be performed will be based on rewriting
the above $ \epsilon $-expansions
in such a way that they may be considered
as strong-coupling expansion of functions, whose weak-coupling expansions
are provided by power series expansion
in powers of $ \varepsilon =4-D$, which are known from $\phi^4$-theory in  $D=4- \varepsilon $
dimensions.
In this way we shall be able to derive
accurate critical exponents $ \nu ^{-1}$ from the non-Borel expansion
(\ref{@ps}).

~\\{\bf 4.}
Let us briefly recall
the interpolation procedure
\cite{int}
by which
a divergent weak-coupling  expansion in some variable $g_0$
of the type
$ E_N(g_0)=\sum_{n=0}^N  a_n  g_0  ^n$
can be combined with
a strong-coupling expansion
of the type
 $E_M(g_0)=  g_0  ^{p/q}\sum_{m=0}^M b_m  (g_0 ^{-2/q}) ^m$.
Previously treated examples
\cite{int}
were the
anharmonic oscillator with
parameters
$p=1/3,~q=3$
for the energy eigenvalues,
and the  Fr\"ohlich
polaron with $p=1,~q=1$ for the
 ground-state energy
and
$p=4,~q=1$ for the mass.
As described in detail in \cite{PI},
the
first step is
to rewrite the
 weak-coupling expansion
with the help of an auxiliary scale parameter $   \kappa$
as
\begin{equation}
E_N(g_0)  =  \kappa ^p \sum_{n=0}^{N} a_n
     \left(\frac{ g_0  }{ \kappa^q }\right)^n
\label{wcexp}\end{equation}
where $ \kappa $
is
eventually set
 equal to $1$. We shall see below that the quotient $p/q$
parametrizes
the
{\em leading power behavior\/}  in $g_0$ of the strong-coupling expansion,
whereas
$2/q$ characterizes the {\em approach\/} to the leading power behavior.
In a second step we replace  $ \kappa$ by the identical expression
\begin{equation}
  \kappa  \rightarrow  \sqrt{ \Kappa ^2 +  \kappa ^2 - \Kappa ^2 }
\label{kappar}\end{equation}
containing a dummy scaling parameter $ \Kappa$.
The series (\ref{wcexp}) is then reexpanded
in powers of $ g_0 $ up to the order $N$,
thereby treating $ \kappa ^2 - \Kappa ^2  $
 as a quantity of order $ g_0  $.
The result is most conveniently expressed
in terms of dimensionless parameters
 $\hat  g_0   \equiv { g_0  }/
{ \Kappa ^q}$ and $ \sigma\equiv (1-\hat  \kappa^2)/\hat g_0$,
where $\hat  \kappa\equiv  \kappa/ \Kappa$.
Then the
replacement
 (\ref{kappar})
amounts to
\begin{equation}
 \kappa \longrightarrow  \Kappa (1- \sigma \hat g_0)^{1/2} ,
\label{subst}\end{equation}
so that the reexpanded
series reads explicitly
\begin{equation}
 W_{N}(\hat g_0, \sigma ) = \Kappa^p \sum_{n=0}^{N}
 \varepsilon_{n}( \sigma )
\left({\hat g}_0 \right)^n,
\label{wnexp}
\end{equation}
 with the coefficients
\begin{equation}
   \varepsilon_{n}( \sigma ) = \sum_{j=0}^{n}  a_{j}
      \left( \begin{array}{c}
              (p - q j)/2 \\ n-j
             \end{array}
      \right)
      (-\sigma )^{n-j}.  \label{rec}
\end{equation}
For any fixed $g_0$,
we
form the first and second derivatives of
$W_N( g_0  , \Kappa )$ with respect to $ \Kappa $,
calculate the $ \Kappa$-values
of the extrema and the turning points. If there is a unique extremum,
this supplies us
with an optimal scaling parameter $ \Kappa _N$.
If no extremum exists, we use the turning point
to determine $ \Kappa _N$.
If there are more than one
extremum  or turning point, we
take the
smallest of these as $ \Kappa _N$.
This procedure is called  {\em optimization\/}.
The function $ W_N ( g_0 )\equiv  W_N ( g_0  ,  \Kappa _N) $
constitutes the $N$th variational approximation  $E_N(g_0)$
to the function $E(g_0)$.

It is easy to take this
approximation
to the strong-coupling limit $ g_0 \rightarrow \infty$.
For this we observe that
(\ref{wnexp}) has the
 scaling form
\begin{equation}
 W_N ( g_0  , \Kappa ) =  \Kappa ^p
 w_N ( \hat g_0  , \hat  \kappa^2).
\label{scalf}\end{equation}
For dimensional reasons, the optimal
 $ \Kappa _N $ increases with $ g_0  $
like
 $ \Kappa _N\approx  g_0  ^{1/q}c_N  $,
so that $\hat g_0=c_N^{-q}$ and $ \sigma=1/\hat g_0=c_N^{q}$
remain finite in the strong-coupling limit, whereas $\hat  \kappa^2$
goes to zero like
 $1/[c_N( g_0  / \kappa ^q)^{1/q}]^2$.
Hence
\begin{equation}
  W_N ( g_0  ,  \Kappa _N) \approx  g_0  ^{p/q}  c_N^p  w _N(c_N^{-q},0).
\label{gl1}\end{equation}
Here $c_N$ plays the role of the variational parameter
to be determined by applying the optimization process described above
to the function
$c_N^pw_N(c_N^{-q},0)$.
The full strong-coupling expansion is obtained from the
Taylor series of $w_N(\hat g_0, \hat  \kappa^2)$
in powers of $\hat  \kappa^2=( g_0/ \kappa^q\hat g_0)^{-2/q}$, which yields
\begin{eqnarray}
\!\!\!\!\!\! W_N( g_0  )& = & g_0  ^{p/q}
\bigg[~ b_0(\hat g_0) + b_1(\hat g_0) \left(\frac{ g_0  }{ \kappa^q }\right)
               ^{-2/q}
+ b_2(\hat g_0) \left(\frac{ g_0  }{ \kappa^q}\right)^{
                   -4/q} + \dots \bigg]
\label{sc}\end{eqnarray}
with
\begin{equation}
 b_n(\hat g_0) = \left.\frac{1}{n!}  w _N^{(n)} (\hat  g_0  , 0) \hat  g_0  ^{
               (2n-p)/q} \right.
,
\label{detb}\end{equation}
%
where $ w _N^{(n)}(\hat  g_0  , \hat   \kappa^2 )$ is
the $n$th derivatives of $w_N(\hat g_0, \hat   \kappa^2) $
with respect to $\hat  \kappa^2$. Explicitly:
\begin{eqnarray}
&& \frac{1}{n!} w_N^{(n)}(\hat{g}_0,0) =
\sum_{l=0}^N (-1)^{l+n} \sum_{j=0}^{l-n} a_{j}^{\rm }
      \left( \begin{array}{c}
              (p - q j)/2 \\ l-j
             \end{array}
      \right)
      \left( \begin{array}{c}
              l-j \\ n
             \end{array}
      \right)
      (-\hat{g}_0)^j.
\label{coeffb}\end{eqnarray}
The optimal
expansion
of
the energy (\ref{sc}) is obtained by expanding
\begin{equation}
\hat g_0= \gamma _0+
  \gamma _1  \left(\frac{ g_0  }{ \kappa^q }\right) ^{-2/q}
+ \gamma  _2  \left(\frac{ g_0  }{ \kappa^q }\right) ^{-4/q}
+\dots~,~~~~
\label{goh}
\end{equation}
where $\gamma _0=c_N^{-q}$, and finding the
optimal extremum (or turning point)
in the resulting polynomials of $   \gamma  _1,  \gamma  _2,\dots~$.
In this way we obtain a
systematic
strong-coupling
coupling expansion in powers of
$ \left({ g_0  }/{ \kappa^q }\right) ^{-2/q}$.
This is done as follows:
We first optimize the leading strong-coupling
coefficient $b_0(\hat g_0)$ in $\hat g_0$,
and identify the
optimal position
by $ \gamma _0$.
Optimizing
$W_N(g_0)$
with the
expansion (\ref{goh}),
in $ \gamma _1$, $ \gamma _2, \dots~$,
yields for the parameters $p=-2,q=2$ at
the coefficients $ \gamma_1, \gamma _2,\dots~$
and optimal $b_n(\hat g_0)$'s
by the equations listed in
Table \ref{tb1}.

It was demonstrated in \cite{int}
how
one can now find a variational
perturbation series for functions
for which one knows
$N$ weak-coupling  and
$M$ strong-coupling expansion coefficients.
We must merely extend the
set of of  coefficients $a_1,\dots,a_N$
by $M$ unknown ones $a_{N+1},\dots,a_{N+M}$,
and determine the latter via
a fit of
the resulting
strong-coupling coefficients
$b_0,\dots,b_{M-1}$ to the known ones.

~\\
\noindent
{\bf 5.}
This interpolation procedure will now be applied
to the perturbation expansion
(\ref{@ps})
in $2+ \epsilon $ dimensions, considering
it as the {\em strong-coupling expansion\/}
of a series in the variable $\tilde  \varepsilon =2(4-D)/(D-2)
=4(1- \epsilon /2)/ \epsilon = \varepsilon /(1-  \varepsilon /2) $:
\begin{equation}
 \nu ^{-1}(\tilde  \varepsilon )=
 {4}\,{\tilde  \varepsilon^{-1} }
-8 \frac{n-4}{n-2}\,\tilde  \varepsilon^{-2}
+16 \frac{n-4}{n-2}\,\tilde  \varepsilon^{-3 }
-32 \left\{\left[52+108\zeta(3)\right]-\left[16+36\zeta(3)\right]n+n^2 \right\} \,\frac{\tilde  \varepsilon^{-4 }}{(n-2)^3}
+\dots~.
\label{@sce}\end{equation}
The variable $\tilde  \varepsilon $  plays the role of the variable $g_0$ in the general formulas
of the last section.
The weak-coupling
expansion of $ \nu ^{-1}(\tilde  \varepsilon )$ in powers of $\tilde \varepsilon$ can be
obtained
directly from the $\varepsilon$-expansions
 of Ref.~\cite{ksf}, and  has
for $n=3,4,5,1$, the
numerical form
\begin{eqnarray}
 n=3:~~~\nu ^{-1}&=&2 - 0.45455\,\tilde\varepsilon + 0.071375\,{\tilde\varepsilon^2} + 0.15733\,{\tilde\varepsilon^3} - 0.52631\,{\tilde\varepsilon^4} +\dots~,\\
n=4:~~~ \nu ^{-1}&=&
2 - 0.5\,{\tilde \varepsilon} + 0.0833333\,{{{\tilde \varepsilon}}^2} + 0.147522\,{{{\tilde \varepsilon}}^3} -
  0.499944\,{{{\tilde \varepsilon}}^4} + 1.47036\,{{{\tilde \varepsilon}}^5}+\dots~,\\
 n=5:~~~\nu ^{-1}&=&
2 - 0.538462\,{\tilde \varepsilon} + 0.0955849\,{{{\tilde \varepsilon}}^2} +
  0.135442\,{{{\tilde \varepsilon}}^3} - 0.469842\,{{{\tilde \varepsilon}}^4} +
  1.34491\,{{{\tilde \varepsilon}}^5}
 +\dots~,\\
 n=1:~~~\nu ^{-1}&=&
2 - 0.333333\,{\tilde \varepsilon} + 0.0493827\,{{{\tilde \varepsilon}}^2} +
  0.158478\,{{{\tilde \varepsilon}}^3} - 0.539937\,{{{\tilde \varepsilon}}^4} +
  1.78954\,{{{\tilde \varepsilon}}^5}
+\dots~.
\label{@expa}\end{eqnarray}
Extending these series by
four more terms
$a_6\, \tilde\varepsilon^6+a_7\,\tilde\varepsilon^7
+a_8\,\tilde\varepsilon^8
+a_9\,\tilde\varepsilon^9$,
we
calculate the strong-coupling coefficients
(\ref{detb})
by
extremizing (\ref{sc}) with (\ref{goh}),
after
identifying $g_0$ with $\tilde  \varepsilon $.
The parameters $(p,q)$ are equal to $(-2,2)$, as follows directly from
a comparison of the strong-coupling
powers
$\tilde  \epsilon ^{p/q}\left(1+\tilde  \varepsilon ^{-2/q}+\dots\right)$
with (\ref{@sce}).
The coefficients
$a_6\,,a_7,\,a_8,\,a_9$
are now determined to make
$b_0(\hat g_0),
b_1(\hat g_0),
b_2(\hat g_0),
b_3(\hat g_0)$
agree with
(\ref{@sce}).
The technique of doing this is described in detail in
Ref.~\cite{int}.

\begin{figure}[tbhp]
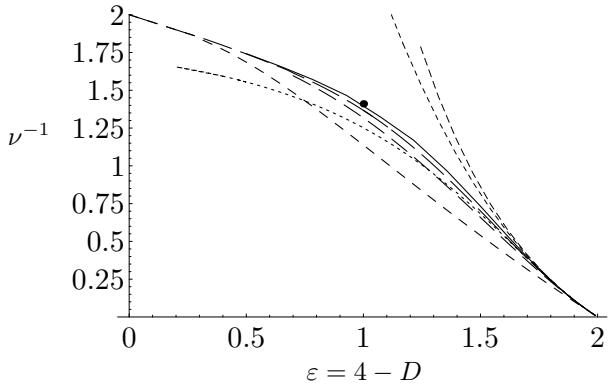

\hspace{-3.cm}\input plnuiff.tps   ~\\
\caption[]{
Inverse of the critical exponent
$ \nu$ for the classical Heisenberg model in the
O(3)-universality class.
Solid curve represents the interpolation result of fourth order.
Lower dashed  curves show interpolations
of first, second, and third order.
Upper short-dashed curves display, with decreasing dash length,
the first three and four terms of
the
$ \epsilon $-expansion (\ref{@ps2}), respectively.
 Dotted curve is
Pad\'e [1,2]-Borel approximations
The fat dot corresponds to
the seven-loop result
in $D=3$ dimensions,
$ \nu =0.7073$ of Ref.~\cite{gz,seven}.
The four
interpolations give
$ (\nu_1, \nu _2, \nu _3, \nu _4)=(
  0.87917,\,
      0.75899,\,
      0.731431,\,0.712152)$.
These are extrapolated
in Fig.~\ref{shpl} to infinite order,
yielding
$ \nu =0.695$.
}
\label{plnuif}\end{figure}

\begin{figure}[tbhp]
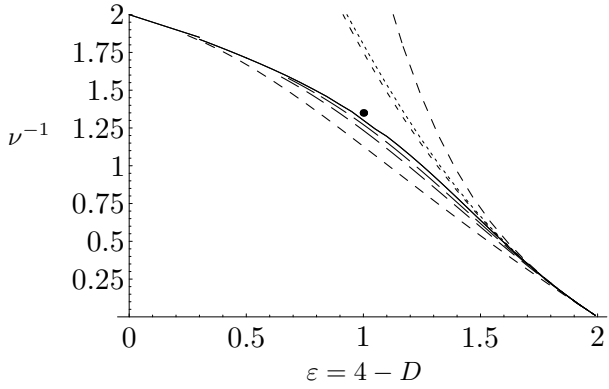

\hspace{-3.cm}\input plnuif4.tps   ~\\
\caption[]{
Same plot as in Fig.~\ref{plnuif}, but for the
O(4)-universality class.
Fat dot represents
six-loop result
in $D=3$ dimensions
$ \nu =0.737$  of Ref.~\cite{SC}.
The four interpolations give
$ (\nu_1, \nu _2, \nu _3, \nu _4)=
(0.88635,\,$$0.810441,\,$$0.786099,\,$$0.768565)$.
The extrapolation
to infinite order
shown in Fig.~\ref{shpl4}
yields
$ \nu =0.735$.
}
\label{plnuif4}\end{figure}

\begin{figure}[tbhp]
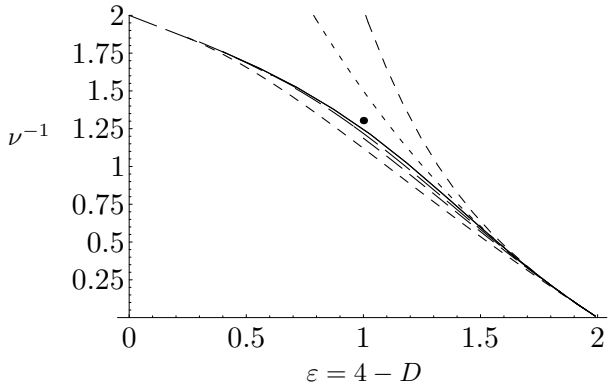

\hspace{-3.cm}\input plnuif5.tps   ~\\
\caption[]{
Same plot as in Fig.~\ref{plnuif}, but for the
O(5)-universality class.
 There exists no
Pad\'e-Borel approximation.
Fat dot represents
ix-loop result
in $D=3$ dimensions
$ \nu =0.767$  of Ref.~\cite{SC}.
The four interpolations give
$ (\nu_1, \nu _2, \nu _3, \nu _4)=
  (0.89278,\,0.842391,\,0.820491,\,0.802416)$.
The extrapolation
to infinite order
shown in Fig.~\ref{shpl5}
yields
$ \nu =0.766$.
}
\label{plnuif5}\end{figure}

In order to see how the result improves with the number
$M$ of additional terms in (\ref{goh}),
we go through this procedure successively
for $M=1,2,3,4$.
The successive additional expansion coefficients
for the  O($n$) universality classes
with $n=3,\,4,\,5,\,1$
are listed in Tables
\ref{tabn3}--\ref{tabn1}, respectively.
The four resulting curves for $~\nu ^{-1}( \varepsilon )$ are
shown in
Figs.~\ref{plnuif}--\ref{plnuif5}.
For $n=3$,
the
successive critical exponents $ \nu $ at $ \varepsilon =1$
taken from Fig.~\ref{plnuif} are $ (\nu_1, \nu _2, \nu _3, \nu _4)=(
  0.87917,\,
      0.75899,\,
      0.731431,\,0.712152)$.
Their $M$-dependence
is plotted in Fig.~\ref{shpl}
as a function of the variable $x=M^{-1.8}$
which makes them lie
approximately on a smooth parabola
intercepting
the $ \nu $-axis at $ \nu _\infty=0.695\pm0.010$.
This extrapolated value
is in good agreement with the
above-quoted value $\approx 0.705$
from  seven-loop calculations in $4- \varepsilon $ dimensions \cite{gz,seven}.
The results for the other O($4$) and O($5$) universality classes
are displayed analogously. The respective $ \nu $-values
$0.735\pm0.010$ and $0.766\pm0.010$ agree well
with the highest available
six-loops results of Ref.~\cite{SC}, which are
$0.737$ and $0.767$.

\begin{figure}[tbhp]
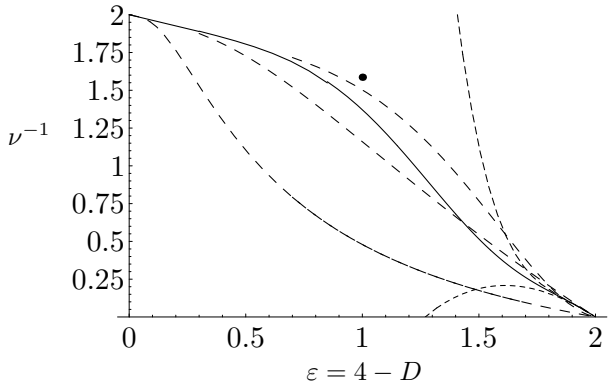

\hspace{-3.cm}\input plnuif1a.tps   ~\\
\caption[]{
Same plot as in Fig.~\ref{plnuif}, but for the
O(1)-universality class (of the Ising model).
 Again there is no
Pad\'e-Borel approximation.
Fat dot represents
seven-loop result
in $D=3$ dimensions
$ \nu =0.6305$  of Refs.~\cite{gz,seven}.
The four interpolations give
$ (\nu_1, \nu _2, \nu _3, \nu _4)=
  (0.862357,\,0.665451,\,2.08686,\,0.729231)$.
Their failure to converge is illustrated
graphically
in Fig.~\ref{shpl5}.
}
\label{plnuif1}\end{figure}

\begin{figure}[tbhp]
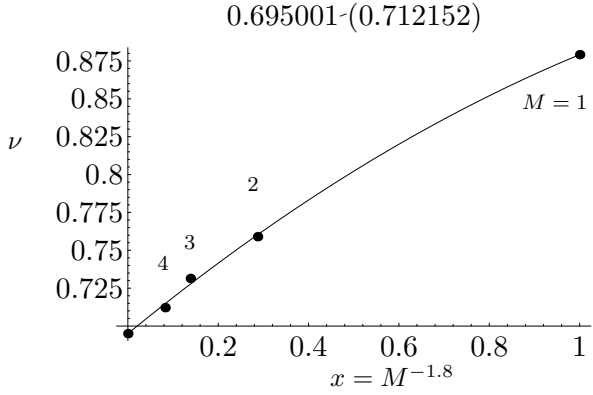

\hspace{-3.5cm}
\input plnuf3.tps   ~\\
\caption[]{
The
four successive approximations
$ (\nu_1, \nu _2, \nu _3, \nu _4)=(
  0.87917,\,
      0.75899,\,
      0.731431,\,0.712152)$
for $n=3$ (Heisenberg model) plotted as a function of
 $x=M^{-1.8}$
which makes them lie
a smooth parabola line
with the intercept
 $ \nu _\infty=0.695\pm0.010$.
Numbers on top show extrapolated value
and highest approximation (in parentheses).
}
\label{shpl}\end{figure}

As discussed above,
the relation between the
$ \epsilon $- and $ \varepsilon $-expansions is
expected to be restricted to $n>2$, for physical reasons.
It is instructive to see
 that the
variational interpolation method
reflects this problem at two places.
First,
the
expansion coefficients in Table \ref{tabn1}
shows a large irregularity for $n=1$.
Second,
the successive approximations
for $ \nu^{-1}$ in Fig.~\ref{plnuif1}
display no tendency of convergence
with increasing order $M$ of approximation.

Finally, we plot our highest ($M=4$) approximations
for $n=3,\,4,\,5$
together with the large-$n$ approximations
for $n=\infty,\,20,\,10,\,6$ in Fig.~\ref{1ovn}
to see the change of the
$\tilde  \varepsilon $-behavior for increasing $n$,
which shows that the latter for $n=6$ is still far from the exact
curve.
This can also be seen in
Fig.~3 of Ref.~\cite{SC}.

For the critical exponent $ \eta $,
the series (\ref{@ps2}) reads
in the variable
$\tilde  \varepsilon $:
\begin{eqnarray}
 &&\eta(\tilde  \varepsilon ) =
 {{\tilde  \varepsilon^{-1} }\over{n-2} }
-2n \frac{\,\tilde  \varepsilon^{-2}}{(n-2)^2}
+8n(n-1) \frac{\tilde  \varepsilon^{-3 }}{(n-2)^3}
+16(n-1)
\left[\left(6-2n^2\right)\left[12-n-n^2\right) \zeta(3) \right]
\frac{\tilde  \varepsilon^{-4 }}{(n-2)^4}
+\dots~,
\label{@scet}\end{eqnarray}
whereas the weak-coupling
expansion
in powers of $\tilde \varepsilon$
obtained
 from the $\varepsilon$-expansions
 of Ref.~\cite{ksf}  has,
%
\cm{{\scriptsize
\begin{eqnarray}
 &&\!\!\!\!\! \frac{\eta}{\epsilon ^2 }= \frac{2+n}{2(n+8)^2}+
{\left( 272 + 56\,n - {n^2} \right) }
\frac{\left( 2 + n \right) \, \epsilon }{8(n+8)^4}
 \nonumber \\
&&~~+\left\{
 46144 + 17920\,n + 1124\,{n^2} -
         230\,{n^3} - 5\,{n^4}
-32\cdot 12\,
\left( 22 + 5\,n \right)
 \,
    {\rm \zeta}(3)\right\}
\frac{\left( 2 + n \right) \, \epsilon^2 }{32(n+8)^6}
  \nonumber \\
&&~~+
\left\{
 5655552 + 2912768\,n + 262528\,{n^2} -
         121472\,{n^3} - 27620\,{n^4} - 946\,{n^5} - 13\,{n^6}
 \right.\nonumber \\
&&~~\left.
+  16 (n+8)\,\left(
 -171264 - 68672\,n - 1136\,{n^2} +
         1220\,{n^3} + 10\,{n^4} + {n^5}
\right)
    {\rm \zeta}(3)
\right.
 \nonumber \\
&&~~\left. +
128\cdot 9\,(n+8)^3\,\left( 22 + 5\,n \right)\,\zeta (4)
 +128\cdot 40\, (n+8)^2\left(
 186 + 55\,n + 2\,{n^2}  \right) \,{\rm \zeta}(5)\right\}
\frac{ \left( 2 + n \right) \,\epsilon^3 }{128(n+8)^8} +\dots~.
\label{@}\end{eqnarray}
}   }
for $N=3,4,5,1$,
the numerical form
\begin{eqnarray}
 N=3:~~~ \eta / \epsilon ^2&=&
{5/ {242}} ~+ 0.0183987\,{\tilde\epsilon} - 0.0166488\,{{{\tilde\epsilon^{2}}}} +
  0.032432\,{{{\tilde\epsilon^{3}}}}+\dots~,\label{@expa3}\\
N=4:~~~  \eta / \epsilon ^2&=&
{1/ {48}} ~+ 0.0173611\,{\tilde\epsilon} - 0.0157657\,{{{\tilde\epsilon^{2}}}} +
  0.029057\,{{{\tilde\epsilon^{3}}}}
+\dots~,\\
 N=5:~~~ \eta / \epsilon ^2&=&
{7/ {338}} ~+ 0.0161453\,{\tilde\epsilon} - 0.0148734\,{{{\tilde\epsilon^{2}}}} +
  0.0259628\,{{{\tilde\epsilon^{3}}}}
 +\dots~,\\
 N=1:~~~ \eta / \epsilon ^2&=&
{1/ {54}} ~+ 0.01869\,{\tilde\epsilon} - 0.0176738\,{{{\tilde\epsilon^{2}}}} +
  0.0386577\,{{{\tilde\epsilon^{3}}}}
+\dots~.
\label{@expa1}\end{eqnarray}
These series can again be extended by
four more terms
$a_4\, \tilde\varepsilon^4+a_5\,\tilde\varepsilon^5
+a_6\,\tilde\varepsilon^6
+a_7\,\tilde\varepsilon^7$,
making the strong-coupling coefficients
$\bar b_0,
\bar b_1,
\bar b_2,
\bar b_3,$
in Eq.~(\ref{detb}) calculated for Eqs.~(\ref{@expa3})--(\ref{@expa1})
agree with those of the expansion of $ \eta /\epsilon^2=\eta/(2-\tilde\varepsilon)^2$
obtained from
Eq.~(\ref{@scet}).
Here, however, we encounter problems:
The $  \eta $-values from the interpolation come out too
large by about a factor $2$.
Also $ \gamma $ does not interpolate well.
A more convenient combination of critical
exponents will have to be found to apply this method.

~~\\~~\\
{\bf Acknowledgment}~\\~\\
The author thanks
Prof. John Gracey
for extremely helpful communications,
and Drs. A. Pelster, A. Schakel, and V. Schulte-Frohlinde for
several useful discussions.

%
%
~\\[-1.5cm]

%
%
%

\begin{table}[tbhp]
\caption[]{Equations determining the coefficients $b_n(\hat g_0)$
in the strong-coupling expansion (\ref{sc}) and the associated
$ \gamma _i\equiv c_n^{-q}  \delta _i$ in (\ref{goh})
from
the functions $\bar b_n\equiv b_n( \gamma _0)$
and their derivatives. For brevity, we have
suppressed the argument
$\gamma_0$ in these functions.}
\begin{tabular}{cll}
$n$ & $b_n$ & $- \gamma _{n-1} $  \\
\hline
2 &$ \bar b_2+ \gamma _1\bar b'_1+ \frac{1}{2}\gamma _1^2\bar b''_0$ & $\bar b'_1/ \bar b''_0$ \\
3 &$ \bar b_3+ \gamma _2\bar b_1'+ \gamma _1\bar b_2'+
\gamma _1 \gamma _2\bar b_0''+ \frac{1}{2}\gamma _1^2\bar b_1''
+\frac{1}{6} \gamma _1^3\bar b_0^{(3)}$ & $
(\bar b_2'+ \gamma _1\bar b''_1+\frac{1}{2} \gamma _1^2 \bar b_0^{(3)})/\bar b_0''$\\
4 &$ \bar b_4+ \gamma _3\bar b_1'
+ \gamma _2\bar b_2'
+ \gamma _1\bar b_3'+
(\frac{1}{2} \gamma _2^2+\gamma _1 \gamma _3)\bar b_0''$
& $
(\bar b_3'+ \gamma _2\bar b''_1+ \gamma _1 \bar b_2''+ \gamma _ 1\gamma _2\bar b_0^{(3)}$
\\  &
$+ \gamma _1 \gamma _2\bar b_1''
+\frac{1}{2} \gamma _1^2\bar b_2''
+\frac{1}{2}  \gamma _1^2 \gamma _2\bar b_0^{(3)}
+\frac{1}{6}  \gamma _1^3\bar b_1^{(3)}
+\frac{1}{24}  \gamma _1^4\bar b_0^{(4)}
$&$
+\frac{1}{2} \gamma _1^2\bar b_1^{(3)}
+\frac{1}{6} \gamma _1^3\bar b_0^{(4)})/\bar b_0''
$
\end{tabular}
\label{tb1}\end{table}

\begin{table}[tbhp]
\caption[]{Coefficients of
the successive extension of the expansion coefficients in Eq.~(\ref{@expa})
for $n=3$
determined from $M=1,\,2,\,3,\,4$ strong-coupling coefficients
$(4,\,8,\,-16,\,160)$
of Eq.~(\ref{@sce}).
}
\begin{tabular}{cllll}
$n$
& $a_6$ & $a_7$ & $a_8$ & $a_9$ \\
\hline
1&$-203.827$&&&\\
2&$-5.67653$&$17.6165$&&\\
3&$-4.25622$&$9.04109$&$-15.7331$&\\
4&$-3.80331$&$6.87304$&$-10.0012$&$12.3552$
\end{tabular}
\label{tabn3}\end{table}

\begin{table}[tbhp]
\caption[]{Coefficients of
the successive extension of the expansion coefficients in Eq.~(\ref{@expa})
for $n=4$
determined from $M=1,\,2,\,3,\,4$ strong-coupling coefficients
$(4,\,0,\,0,\,221.096)$
of Eq.~(\ref{@sce}).
}
\begin{tabular}{cllll}
$n$
& $a_6$ & $a_7$ & $a_8$ & $a_9$ \\
\hline
1&$-147.508$&&&\\
2&$-7.91064$&$37.1745$&&\\
3&$-4.59388$&$12.3044$&$-27.0837$&\\
4&$-3.72613$&$7.47851$&$-12.2129$&$16.9547$
\end{tabular}
\label{tabn4}\end{table}

\begin{table}[tbhp]
\caption[]{Coefficients of
the successive extension of the expansion coefficients in Eq.~(\ref{@expa})
for $n=5$
determined from $M=1,\,2,\,3,\,4$ strong-coupling coefficients
$(8,\,-8/3,\, 16/3,\, 106.131)$
of Eq.~(\ref{@sce}).
}
\begin{tabular}{cllll}
$n$
& $a_6$ & $a_7$ & $a_8$ & $a_9$ \\
\hline
\hline
1&$-108.648$&&&\\
2&$-10.1408$&$60.7217$&&\\
3&$-4.75598$&$15.1045$&$-38.9689$&\\
4&$-3.57909$&$7.84272$&$-14.1142$&$21.6045$
\end{tabular}
\label{tabn5}\end{table}

\begin{table}[tbhp]
\caption[]{Coefficients of
the successive extension of the expansion coefficients in Eq.~(\ref{@expa})
for $n=1$
determined from $M=1,\,2,\,3,\,4$ strong-coupling coefficients
$(4,\,-24,\,48,\,3825.54)$
of Eq.~(\ref{@sce}).
}
\begin{tabular}{cllll}
$n$
& $a_6$ & $a_7$ & $a_8$ & $a_9$ \\
\hline
1&$-413.921$&&&\\
2&$-5.25285$&$12.1104$&&\\
3&$-442759$&$12450066$&$-196950675$&\\
4&$-5.7343$&$13.7134$&$-25.226$&$38.0976$
\end{tabular}
\label{tabn1}\end{table}

\clearpage

\begin{figure}[tbhp]
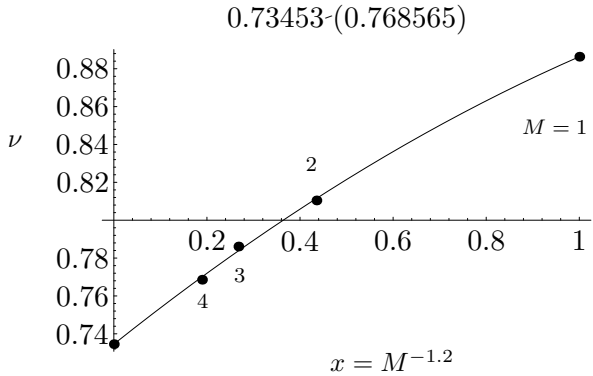

\hspace{-3.5cm}
\input plnuf4.tps   ~\\
\caption[]{
The
four successive approximations
$ (\nu_1, \nu _2, \nu _3, \nu _4)=
(0.88635,\,0.810441,\,0.786099\,0.768565)$
for $n=4$ plotted as a function of
 $x=M^{-1.2}$
which puts them
on a smooth parabola
with the intercept
 $ \nu _\infty=0.735\pm0.010$.
Numbers on top show extrapolated value
and highest approximation (in parentheses).
}
\label{shpl4}\end{figure}

\begin{figure}[tbhp]
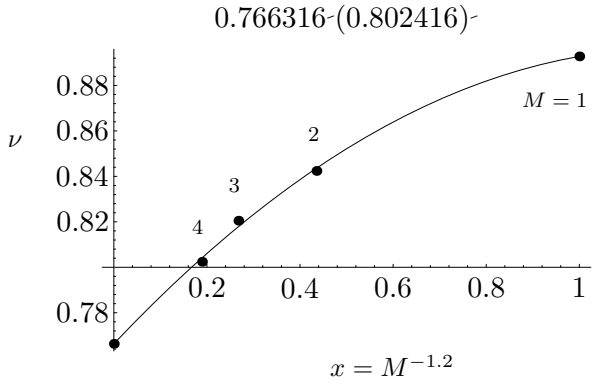

\hspace{-3.5cm}
\input plnuf5.tps   ~\\
\caption[]{
The
four successive approximations
$ (\nu_1, \nu _2, \nu _3, \nu _4)=(0.89278,\,0.842391,\,0.820491,\,0.802416)$
for $n=5$ plotted as a function of
 $x=M^{-1.2}$
which puts them
on a smooth parabola
with intercept
 $ \nu _\infty=0.766\pm0.010$.
Numbers on top show extrapolated value
and highest approximation (in parentheses).
}
\label{shpl5}\end{figure}

\begin{figure}[tbhp]
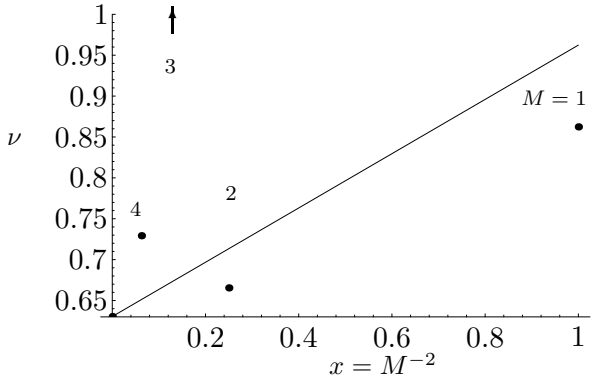

\hspace{-3.5cm}\input shpl1.tps   ~\\
\caption[]{
The
four successive approximations
$ (\nu_1, \nu _2, \nu _3, \nu _4)=
  (0.862357,\,0.665451,\,2.08686,\,0.729231)$
for $n=1$ plotted as a function of
 $x=M^{-2}$.
They show no tendency of convergence
towards the known seven-loop exponent
 $ \nu _\infty=0.630$.
}
\label{shpl1}\end{figure}

\begin{figure}[tbhp]
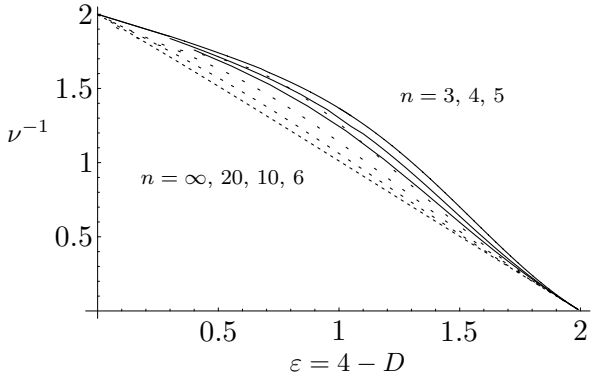

\hspace{-3.5cm}\input pl1ovna.tps   ~\\
\caption[]{
Comparison of $ \nu ^{-1}$ from the highest approximations of our interpolating
 resummation for the O($n$) universality classes with $n=3,4,5$
(counting from the top), with the values obtained
from the $1/n$-expansion to order $1/n^2$ for
$n=\infty,20,10,6$ (counting from the bottom).
The $n=6$ curve is still far from the exact one.
}
\label{1ovn}\end{figure}

\end{document}